\documentclass[twocolumn,amsmath,amssymb,aps,]{revtex4-1}
\usepackage{bm}
\usepackage[colorlinks=true,urlcolor=blue,linkcolor=blue,citecolor=blue]{hyperref}
\usepackage{color}
\usepackage{graphics}
\usepackage{graphicx}
\usepackage{epsfig}
\usepackage{amssymb}
\usepackage{amsmath}
\usepackage{hyperref}
\usepackage{physics}

\usepackage{array}                 

\usepackage{amsfonts}           

\begin{document}

\title{Enhancing non-destructive mass identification via Fourier-transform fluorescence analysis}

\author{F.~Dom\'inguez$^1$}\email{This is part of the PhD Thesis of F. Dom\'inguez}
\author{D. Yousaf$^1$}
\author{J. Berrocal$^1$}
\author{M.J. Guti\'errez$^{1}$}\email{Present address: University of Greifswald, Germany, GSI Darmstadt, Germany}
\author{J. S\'{a}nchez$^1$}
\author{M. Block$^{2,3,4}$}
\author{D. Rodr\'{\i}guez$^{1,5}$}
\email{danielrodriguez@ugr.es}
\affiliation{
$^1$Departamento de F\'{i}sica At\'{o}mica, Molecular y Nuclear, Universidad de Granada, 18071 Granada, Spain\\
$^2$Department Chemie - Standort TRIGA, Johannes Gutenberg-Universität Mainz, D-55099 Mainz, Germany\\
$^3$GSI Helmholtzzentrum für Schwerionenforschung GmbH, D-64291 Darmstadt, Germany\\
$^4$Helmholtz-Institut Mainz, D-55099 Mainz, Germany\\
$^5$Centro de Investigaci\'{o}n en Tecnolog\'{i}as de la Informaci\'{o}n y las Comunicaciones, Universidad de Granada, 18071 Granada, Spain}

\date{\today}%

\begin{abstract}
Single-ion mass identification is important for atomic and nuclear physics experiments on ions produced with low yields. Cooling the ion to ultra-low temperatures by interacting with a laser-cooled ion will enhance the precision of the measurements. In this paper we present axial-common-mode frequency measurements of balanced and unbalanced Coulomb crystals from the Fourier transform of the fluorescence photons from a Doppler-cooling transition in calcium ions, after probing the ion/crystal with a 5-radiofrequency comb. A single ion non-destructively detected can be used for identification yielding a mass resolving power $m/\Delta m_\mathrm{FWHM}\approx 310$ from the axial common mode. This identification can be performed from a single measurement within times below one second. 
\end{abstract}

\maketitle


Experimental studies of atomic and nuclear observables such as masses, nuclear spins, and electromagnetic nuclear moments are important to understand the peculiar properties of the heaviest elements in the Periodic Table of chemical elements \cite{Smit2023,Smit2024}. These nuclides can only be produced artificially in nuclear reactions at accelerator facilities, with very low resulting yields, on the order of ions per second for nuclides with $Z= 102$ down to about one particle per day for $Z=118$. This calls for single-ion sensitivity for laser spectroscopy or mass spectrometry and also a good control and localization of the superheavy element (SHE) ion. For example, the higher spectral resolution needed to resolve the hyperfine splitting \cite{block2021}, required to determine electromagnetic moments and to enable an unambiguous assignment of the nuclear spin, can be attained when cooling the ion to ultra-low temperatures by means of Doppler laser cooling. Since this is not yet feasible due to the limited experimental information on atomic transitions and the often-complex level structure of the atomic spectra, the SHE ion has to be cooled through another (laser-cooled) ion. The same experimental  platform, i.e., an unbalanced Coulomb crystal made of the laser-cooled and the target ion, can provide non-destructive mass identification of the latter \cite{drewsen2007}, and even identification of a molecular ion after inducing some chemical reaction of the SHE ion with reactive gases \cite{drewsen2004}.

The first demonstration of mass spectrometry with sympathetically-cooled ions was reported in 1996 \cite{baba1996}, which has been shown, in the last decade, beneficial to identify stable ions \cite{schneider2014} as well as radioactive ions \cite{sels2022} when working with large ion clouds or accumulating sufficient statistics, respectively. Single-ion mass spectrometry through Fourier transformation of the fluorescence signal from a laser-cooled stable ion was reported in 2001 \cite{schlemmer2001}. Well-localized atomic and molecular ions in Coulomb crystals were observed around the same time \cite{bowe1999,molhave2000}, followed by the demonstration of non-destructive identification through fluorescence images of balanced and unbalanced  two-ion Coulomb crystals \cite{drewsen2004}. In the latter case, the two ions with mass-to-charge ratios $m_s/q_s$ and $m_t/q_t$, have sufficiently small kinetic energy to perform small oscillations around their equilibrium positions. In a linear Paul trap with $\omega_z < \omega_{x,y}$, the ions will line up along the $z$ axis. Considering that the ions move with small oscillation amplitudes compared to their separation distance $D$ and defining $\mu = m_t / m_s$ and $\kappa =q_t/q_s$, the axial motion of the system can be described in terms of normal modes, with eigenfrequencies \cite{gutierrez2021}
\begin{equation}
\begin{split}
\left(\Omega_{z}^{\pm}\right)^2 & \\
=& \frac{\omega_{z}^2}{2} \, \left \lbrace\left [\alpha+\beta\frac{2\kappa}{\kappa+1} \right]\pm \sqrt{\left [\alpha+\beta\frac{2\kappa}{\kappa+1} \right]^2 -12\frac{\kappa}{\mu}} \right \rbrace,
\label{eq:axial_freq_crystal}
\end{split}
\end{equation}
where $\omega_{z}$ is the axial frequency of the single laser-cooled ion ($m_s/q_s$), $\alpha =1 + \kappa/\mu $, and $\beta =1 + 1/\mu $. The superscripts $-$ and $+$ represent the common and the stretch modes, respectively. 
When both ions are singly charged ($q_s=q_t=1$), as it is the case for the experiments presented here, Eq.~(\ref{eq:axial_freq_crystal}) simplifies to the well-known expressions given in Ref.~\cite{morigi2001}. 

Experimentally, if one ion is laser-cooled (e.g. the one with mass-to-charge ratio $m_s/q_s$), the other one will be sympathetically-cooled due to the Coulomb interaction \cite{larson1986}.

Several identification methods have been developed so far for ion crystals in Paul traps: i) resonant excitation of the common mode with a time-oscillating voltage \cite{staanum2008}, ii) application of periodic voltage pulses to one of the trap electrodes \cite{sheridan2011}, iii) scan of the motional frequencies $\Omega_{z}^{\pm}$ through sideband spectroscopy on an electric quadrupole transition with an ultra-stable laser \cite{goeders2013}, iv) determination of the shift in the equilibrium position of the fluorescing ion after the crystal is formed \cite{groot2019}, v) optical amplification of the crystal's motion with lasers \cite{fan2021}, and vi) fast switching of the trap center position with a voltage pulse \cite{saito2022}. The methods i) and iv) are based on the monitoring of the fluorescence signal with an electron-multiplying charge-coupled device (EMCCD) camera, whereas the methods ii), v) and vi) rely on the Fourier analysis of the fluorescence signal detected with a photomultiplier tube (PMT). In this work we detect and identify singly-charged atomic ions through the Fourier analysis of the detected fluorescence photons scattered by the laser-cooled ion in an unbalanced two-ion crystal, after probing the crystal with a radiofrequency comb. The foundation relies on the fluorescence modulation due to the crystal's motion in the trap under near-resonant oscillating electric fields and its subsequent Fourier transform to obtain a discrete frequency spectrum around the axial-common mode. The method is fast, non-destructive, and thus suitable to perform experiments where the highest sensitivity is requested. This is the case of SHEs \cite{block2010,minaya2012,kaleja2022}, provided the full process before the measurement, which comprises on-line production, stopping in a gas cell, separation and subsequent cooling, is carried out with the highest efficiency. We analyze the precision in mass identification obtained with this method in the Doppler regime and discuss the effect of non-linearities due to the Coulomb interaction.

\begin{figure}[t]
\centering
\includegraphics[width=0.48\textwidth]{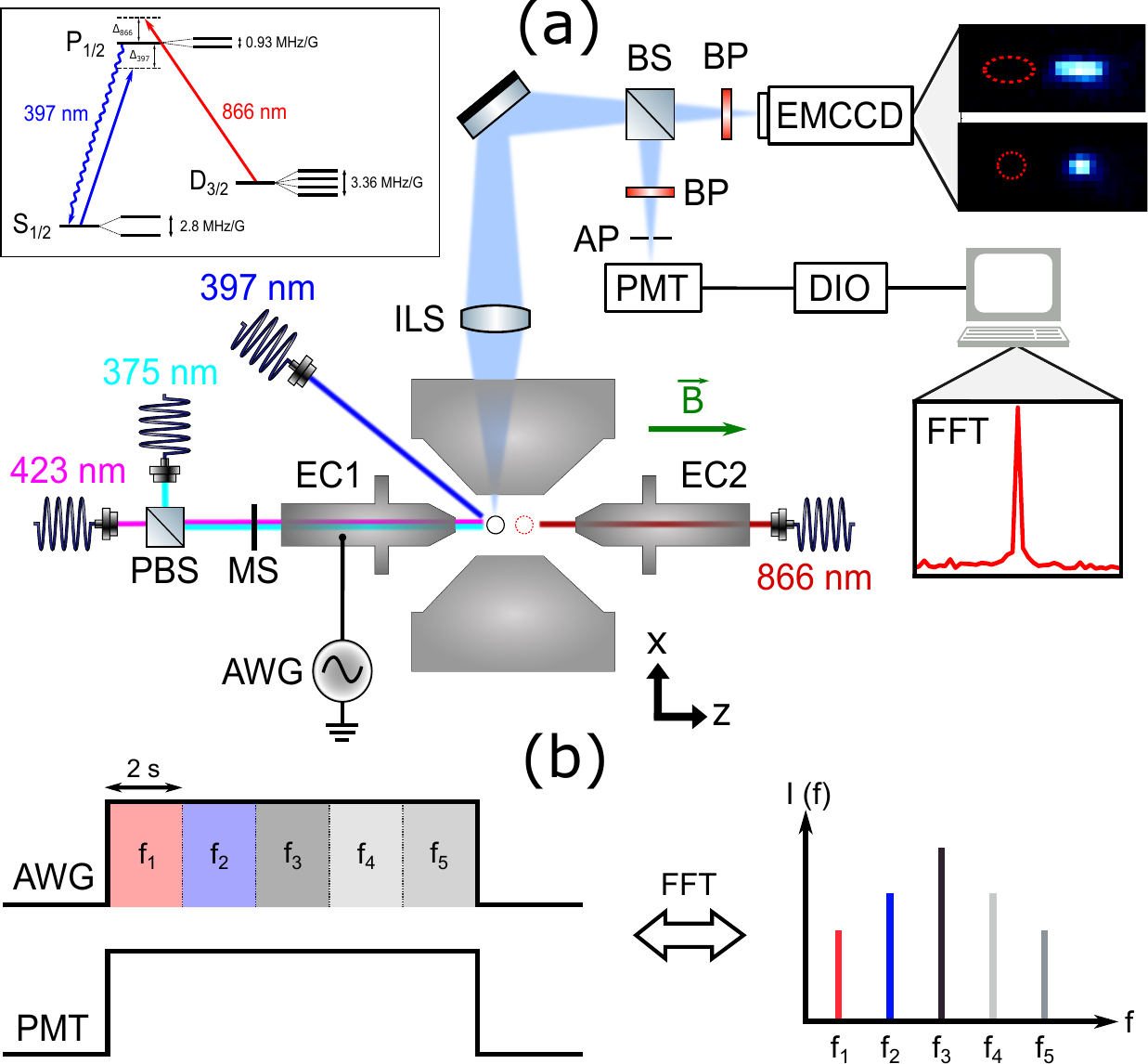}
\vspace{-0.5cm}
\caption{(a) Sketch of the experimental setup for production, Doppler cooling, and identification of $^{A}$Ca$^+$ - $^{40}$Ca$^+$ crystals. The energy levels and transitions employed for Doppler cooling and detection of Ca$^+$ are shown in the inset. The separation distance between the dark (dashed-red circle) and bright ion is $D\approx 8$~$\mu$m. (b) Variation of the excitation-field radiofrequency around resonance and expected frequency spectrum from the PMT detected signal.}
\label{fig1}
\end{figure}

The experimental setup is schematically shown in Fig.~\ref{fig1}(a). A sketch of the radiofrequency comb is also depicted in Fig.~\ref{fig1}(b). The experiments presented here were conducted with two-ion crystals comprised of a Doppler cooled $^{40}$Ca$^+$ ion and a $^{A}$Ca$^+$ isotope ($A$~=~${40, 42, 44, 48}$). For Doppler cooling and detection of $^{40}$Ca$^+$, a 10-MHz red-detuned 397-nm diode laser driving the ${\mathrm{S}_{1/2} \leftrightarrow \mathrm{P}_{1/2}}$ transition was used. A tunable 866-nm diode laser was tuned to the ${\mathrm{D}_{3/2} \leftrightarrow \mathrm{P}_{1/2}}$ repumping transition to decouple the metastable $\mathrm{D}_{3/2}$ state from the cooling cycle. The trap consists of two opposite pairs of blade-shaped electrodes and a pair of endcap-shaped electrodes, with characteristic distances of $2r_0 = 1.6$~mm and $2z_0 = 5.5$~mm for the blade and endcap electrodes, respectively \cite{berrocal2018} (based on Ref.~\cite{Hempel2014}). Radial confinement in the $xy$ plane is accomplished by driving a pair of opposite blades with a radiofrequency (RF) field with ${\omega_{\mathrm{RF}}= 2\pi \times 21.9}$~MHz while keeping the other pair at ground. Axial confinement along the $z$ axis is achieved by applying a DC voltage $U$ to the endcap electrodes (EC1 and EC2 in Fig.~\ref{fig1}). The frequency $\omega _z$ is proportional to $\sqrt{U}$. The trapping voltages are set such that the oscillation frequencies of a single trapped  $^{40}$Ca$^{+}$ ion are $\omega_{x,y} \simeq 2\pi \times1.5$~MHz and $\omega_z \simeq 2\pi \times 0.6$~MHz. A dipolar electric field with amplitude $V_{\hbox{\scriptsize{dip}}}$ can be applied between EC1 and EC2 by means of an arbitrary function generator (AWG), varying the radiofrequency following the comb-like pattern of Fig.~\ref{fig1}(b). The Ca$^+$ ions were produced near the trap center through two-step photoionization of Ca atoms vaporized from a resistively-heated source filled with natural calcium. A free-running diode laser at 375~nm and a tunable diode laser at 423~nm were used. The latter was tuned to the corresponding resonance of the $^1$S$_{0} \leftrightarrow ^1$P$_1$ transition in Ca to selectively produce the ion of interest  \cite{lucas2004}. All tunable lasers were stabilized in frequency using a 10~MHz-accurate wavemeter \cite{dominguez2022}.  The 397-nm fluorescence photons were collected by an imaging lens system (ILS) and split into two branches by a 50:50 beamsplitter (BS) to detect simultaneously the fluorescence signal with an EMCCD camera and a PMT. The EMCCD camera provides spatially-resolved signals that allow certifying the production and characterizing the excitation of calcium crystals (upper right of Fig.~\ref{fig1}). The 397-nm photons were recorded with the PMT for a detection time ${t_{\mathrm{PMT}}}$ and timestamped by a digital input/output (DIO) board. The full experiment is controlled and automated by ARTIQ \cite{artiq}.
Single ions and two-ion crystals were resonantly excited (an example is shown in the upper right side of Fig.~\ref{fig1}) by applying an external dipolar field introducing a modulation of the fluorescence intensity $\delta \mathit{F} (v_z) \propto v_{z,0} \cos (\omega_z \,t)$ such that 

\begin{equation}
\mathit{F} = \Gamma \, \rho_{ee} (v_z) \simeq \Gamma [ \,\rho_{ee,0} + \,\delta \rho_{ee} (v_z)] = \mathit{F}_0 + \delta \mathit{F} (v_z) \; ,
\label{eq:fluo_modulation}
\end{equation}
where $\Gamma$ and $\rho_{ee} (v_z)$ are the spontaneous decay rate and the velocity-dependent occupation probability of the excited state (P$_{1/2}$), respectively. $\rho_{ee,0}$ and $F_0$ are the (approximately) constant occupation probability and fluorescence rate at the Doppler limit. 

\begin{figure}[b]
\centering
\includegraphics[width=0.42\textwidth]{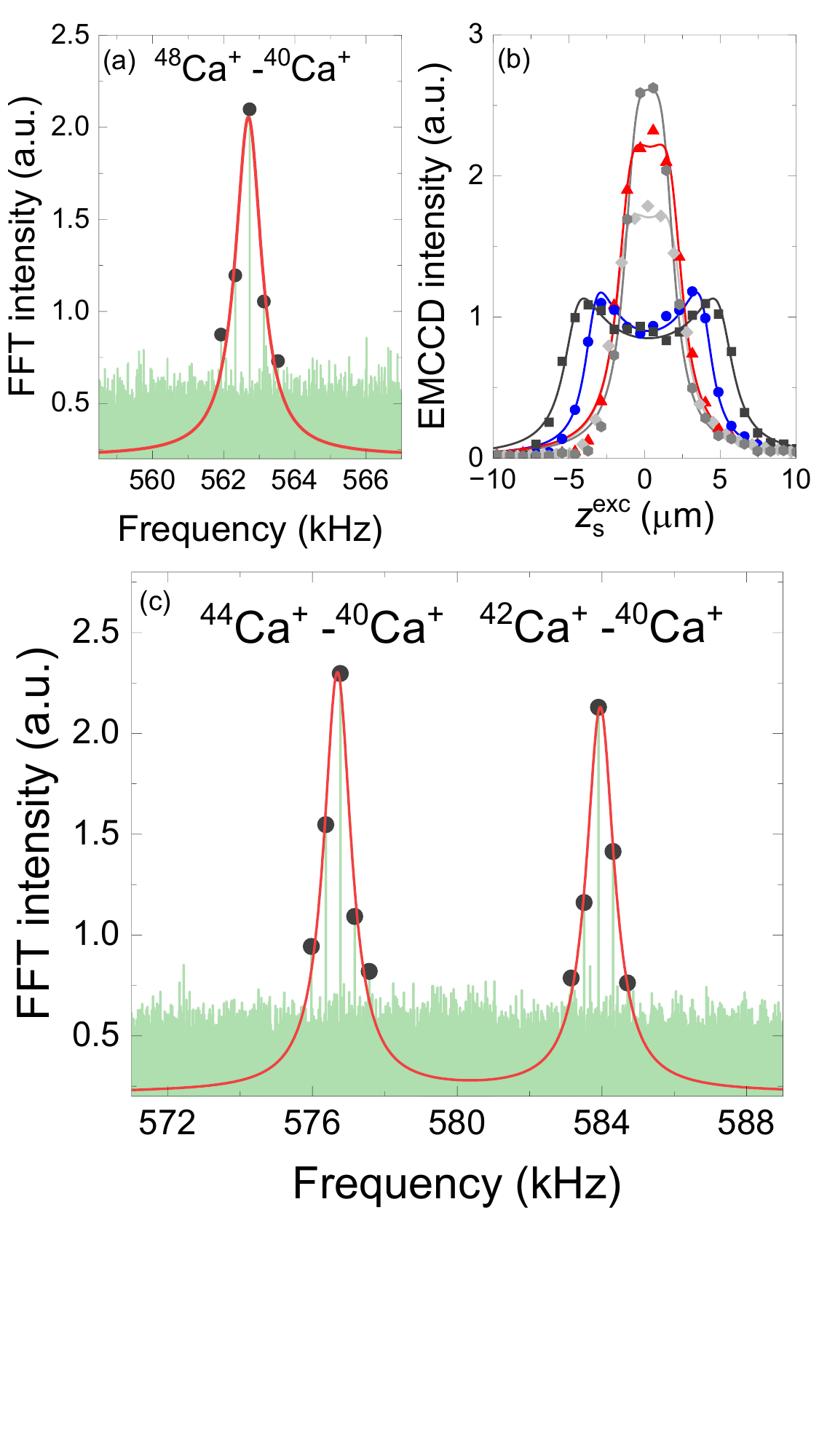}
\vspace{-2.5cm}
\caption{(a) FFT of the detected photons from $f_1$ to $f_5$ of Fig.~\ref{fig1}(b), around $\Omega_{z}^{\scriptsize{\hbox{-}}}$ ($f_3$) for $^{48}$Ca$^+$ - $^{40}$Ca$^+$. The solid circles indicate the FFT intensities for excitations at   $f_1$ to $f_5$. The red-solid curve is a Lorentzian fit considering these data points and the average background level. (b) Axial projections of the EMCCD images when applying the radiofrequency-comb spectrum shown in Fig.~\ref{fig1}(b). The solid lines are fits following the function derived in Ref.~\cite{dominguez2018}. (c) The same as (a) for independent measurements with $^{44}$Ca$^+$~-~$^{40}$Ca$^+$ and $^{42}$Ca$^+$~-~$^{40}$Ca$^+$. Each spectrum is extracted from a single measurement with $t_{\scriptsize{\hbox{w}}}=10$~s.}
\label{fig2}
\end{figure}

\begin{figure}[t]
\centering
\includegraphics[width=0.5\textwidth]{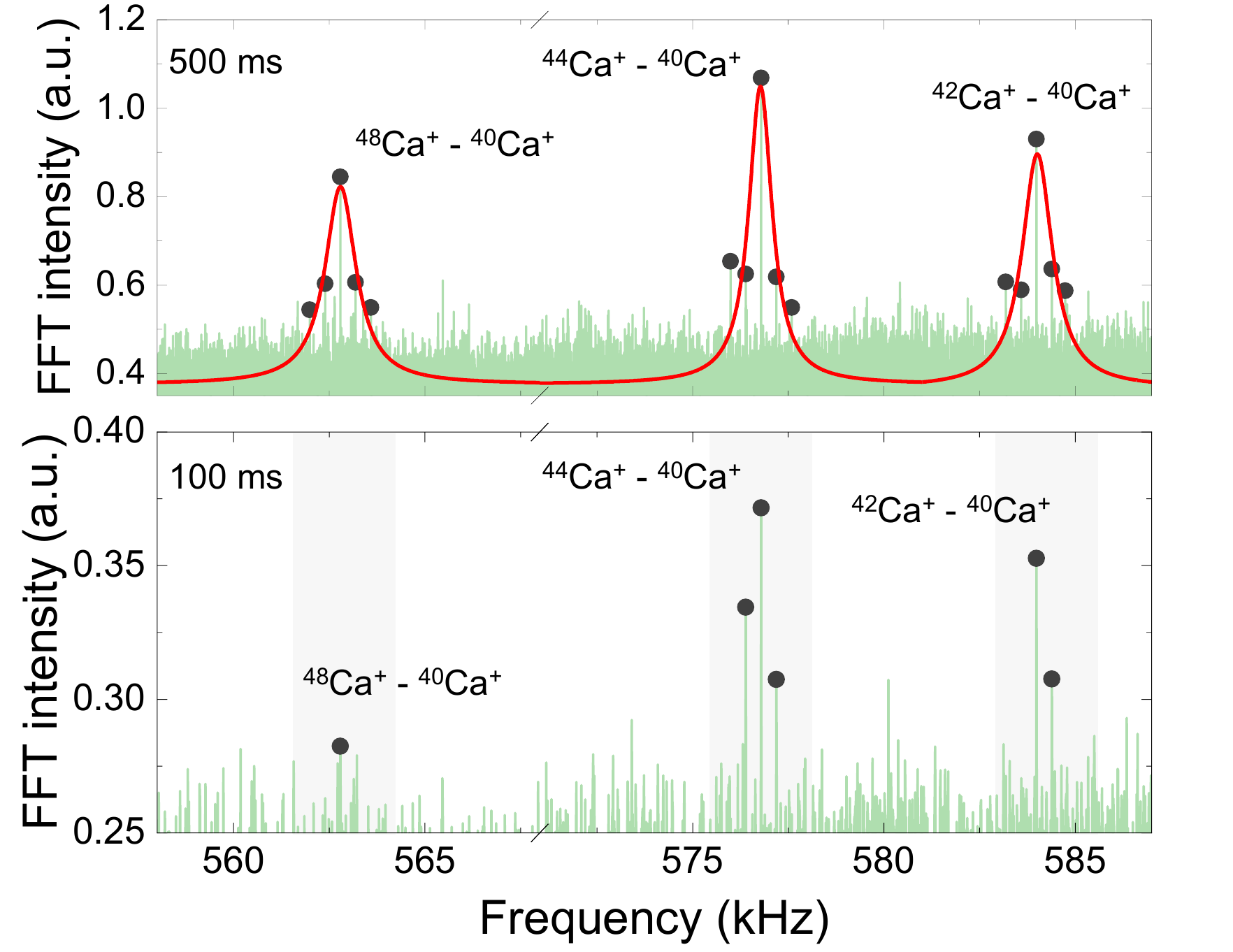}
\vspace{-0.7cm}
\caption{FFT of the detected photons for $^{A}$Ca$^+$ - $^{40}$Ca$^+$ ($A=42,44,48$) from single measurements with $t_{\scriptsize{\hbox{w}}}=500$~ms (top) and  $t_{\scriptsize{\hbox{w}}}=100$~ms (bottom). The amplitude of the dipolar field is 3~mV$_{\hbox{\scriptsize{pp}}}$. The red solid curves are Lorentzian fits to the data. A partial response to the radiofrequency comb is still observed when $t_{\scriptsize{\hbox{w}}}=100$~ms. }
\label{fig3}
\end{figure}

The frequencies $\omega _z$ and $\Omega _z^-$ are determined through the discrete Fast Fourier Transform (FFT) of the detected signal after applying the comb-like radiofrequency field shown in Fig.~\ref{fig1}(b). We analyze the signal recorded during a time window $t_w$ that may be a submultiple of the PMT acquisition time. The data set consists of five intervals, one for each tooth of the radiofrequency comb. The timestamped events were grouped in histograms of bin size $b_{\scriptsize{\hbox{s}}}=250$~ns. The sampling frequency and the frequency resolution are given by $f_{\scriptsize{\hbox{s}}}=1/b_{\scriptsize{\hbox{s}}}$ and $f_{\scriptsize{\hbox{r}}}=1/t_{\scriptsize{\hbox{w}}}$, respectively. 

FFT signals for different unbalanced two-ion crystals are shown in Fig.~\ref{fig2}(a)~and~(c) for an excitation-field amplitude $V_{\hbox{\scriptsize{dip}}}=2$~mV$_{\hbox{\scriptsize{pp}}}$. The fluorescence signal was recorded with the PMT for $t_{\mathrm{PMT}} = 10$~s (equal to $t_{\scriptsize{\hbox{w}}}$ in this case). The signal-to-noise ratio (SNR) has been evaluated for different values of $t_w$ (see Suppl. material). The frequency resolving power $\Omega _z/\Delta \Omega_{z,\mathrm{FWHM}} = 620$ results in a mass resolving power $m/\Delta m_\mathrm{FWHM}\approx 310$. When $t_{\scriptsize{\hbox{w}}}<10$~s, zero padding was applied to obtain the same value of $f_r$. Figure~\ref{fig2}(b) shows the axial oscillation of the sensor ion, obtained from the EMCCD image, for the different data points (comb frequencies) of Fig.~\ref{fig2}(a). The maximum oscillation amplitude ($z_s^{\hbox{\scriptsize{max}}}$) goes from $1.14(2)$ to $4.95(3)$~$\mu$m, for the minimum and maximum FFT intensity, respectively. Figure~\ref{fig3} shows FFT signals from shorter individual measurements with $t_{\scriptsize{\hbox{w}}}=500$~ms (consisting of 5 subintervals of 100~ms) and 100~ms (5 subintervals of 20~ms), and $V_{\hbox{\scriptsize{dip}}}=3$~mV$_{\hbox{\scriptsize{pp}}}$. Featured peaks are always observed even for $t_{\scriptsize{\hbox{w}}}=100$~ms, although fitting curves were obtained only when $t_{\scriptsize{\hbox{w}}}\geq 500$~ms for 3~mV$_{\hbox{\scriptsize{pp}}}$ as shown in Fig.~\ref{fig3}.

Alternating frequency measurements with single Doppler-cooled ions and two-ion crystals have been performed to evaluate the optical mass identification method with the setup of Fig.~\ref{fig1}. A reference measurement of $\omega_{z}$ with a single Doppler-cooled $^{40}$Ca$^+$ ion was carried out before (at time $t_{i-1}$) and after ($t_{i+1}$) a measurement of $\Omega _z^-$ of a \mbox{$^{A}$Ca$^+$ - $^{40}$Ca$^+$} crystal at time $t_i$ to account for fluctuations of the voltage $U$ or any other possible factors that perturb $\omega _z $ and $\Omega _z ^-$. Each measurement in turn consisted of ten consecutive measurements with ${t_{\mathrm{PMT}}}=10$~s each. A linear interpolation between the two reference measurements of $\omega_{z}$ was performed at times $t_{i-1}$ and $t_{i+1}$ to obtain $\omega_{z}$ at $t_i$ and in this way the ratio $\xi (t_i) = \Omega_z^{-} (t_i)/\omega_{z} (t_i)$. Recalling Eq.~(\ref{eq:axial_freq_crystal}),  and considering $q_s=q_t=1$, the frequency ratio (in the linear approximation) can be calculated from known mass values as \cite{morigi2001}
\begin{equation}
\xi = \sqrt{1+\frac{1}{\mu}-\sqrt{1+\frac{1}{\mu ^2}-\frac{1}{\mu}}}\; ,
\label{eq:mass_ratio_crystal}
\end{equation}
with  
\begin{equation}
\delta \xi (t_i)= \xi (t_i) \sqrt{\left[ \delta \Omega_{z}^- (t_i)/\Omega_{z}^- (t_i) \right]^2 + \left[ \delta \omega_{z} (t_i)/\omega_{z} (t_i) \right]^2 }. \label{uncert}
\end{equation}

\begin{figure}[t]
\centering
\includegraphics[width=0.49\textwidth]{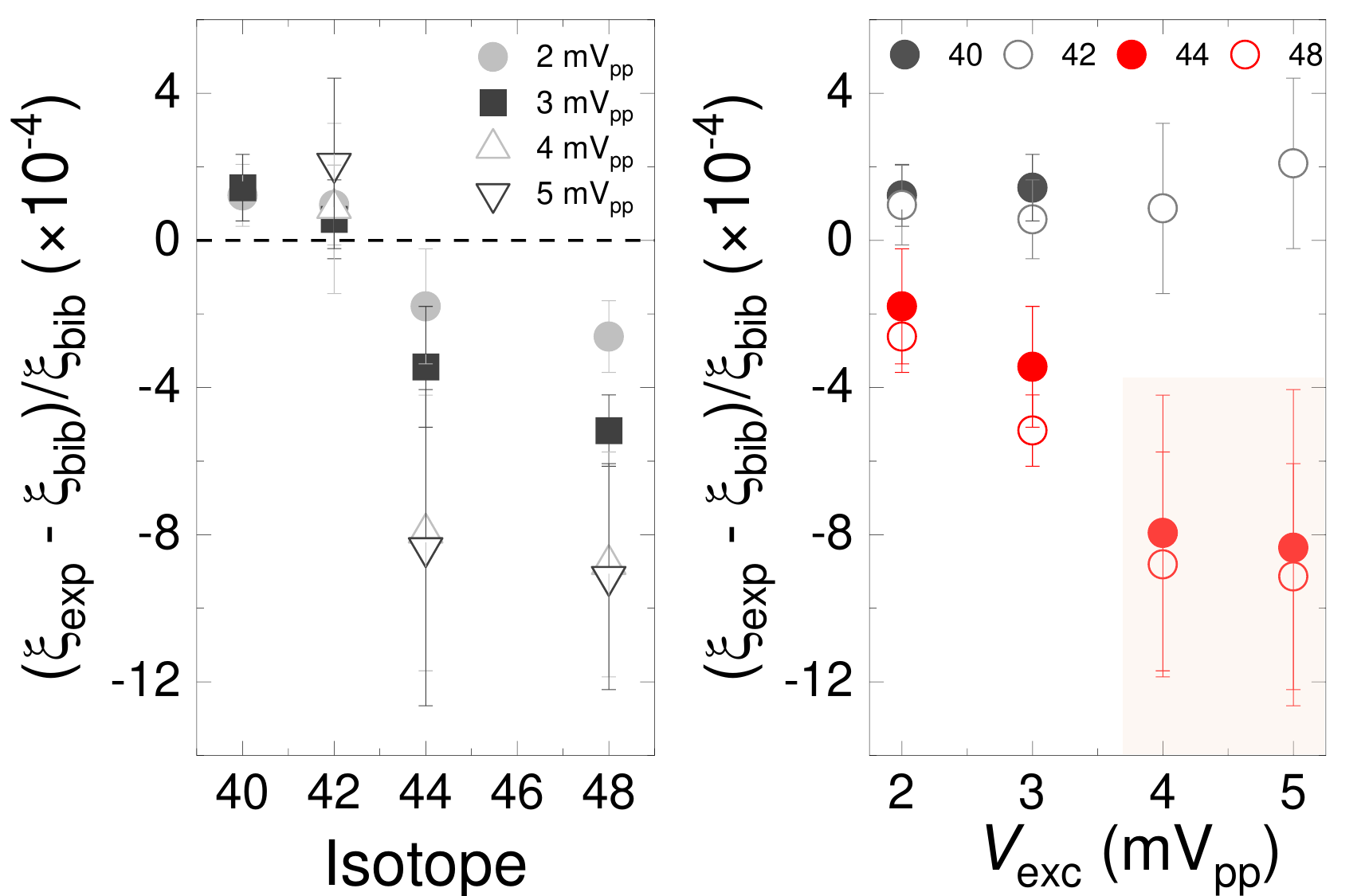}
\vspace{-0.6cm}
\caption{Left: Relative deviation of $\xi _{\hbox{\scriptsize{exp}}}$ with respect to $\xi _{\hbox{\scriptsize{bib}}}$ for $^{A}$Ca$^+$ - $^{40}$Ca$^+$  crystals and different excitation amplitudes. Right: The same but as a function of $V_{\hbox{\scriptsize{dip}}}$. $t_{\scriptsize{\hbox{w}}}=10$~s for these measurements. The same behaviour has been observed when $t_{\scriptsize{\hbox{w}}}=1$~s. When $V_{\hbox{\scriptsize{dip}}}\geq 4$~mV$_{\hbox{\scriptsize{pp}}}$, the ordered structures of the crystals $^{44}$Ca$^+$ - $^{40}$Ca$^+$ and $^{48}$Ca$^+$ - $^{40}$Ca$^+$ vanish for the central frequency of the comb (data points contained in the shaded area of (b)). }
\label{fig4}
\end{figure}

The relative deviations of the measured ratios $\xi _{\hbox{\scriptsize{exp}}}$ ($\xi (t_i)$), with respect to the expected values $\xi _{\hbox{\scriptsize{bib}}}$, for the crystals \mbox{$^{A}$Ca$^+$ - $^{40}$Ca$^+$}, with $m_s/q_s =40$ have been evaluated and are shown in the left side of Fig.~\ref{fig4}. The ratios $\xi _{\hbox{\scriptsize{bib}}}$ are obtained after substituting the values of $\mu$ from the bibliography data \cite{ame2021,berrocal2024} in Eq.~(\ref{eq:mass_ratio_crystal}). The measurements have been carried out for several amplitudes of the excitation field ($V_{\hbox{\scriptsize{dip}}}=2$, 3, 4 and 5~mV$_{\hbox{\scriptsize{pp}}}$). The results for the crystal \mbox{$^{42}$Ca$^+$-$^{40}$Ca$^+$} agree with the expected values within an uncertainty in the order of 10$^{-4}$ in all cases. Under approximately the same experimental conditions, $(\xi _{\hbox{\scriptsize{exp}}}-\xi _{\hbox{\scriptsize{bib}}})/\xi _{\hbox{\scriptsize{bib}}}$ deviates more than one sigma for the crystals $^{44}$Ca$^+$-$^{40}$Ca$^+$ and $^{48}$Ca$^+$-$^{40}$Ca$^+$, and becomes larger when increasing the difference $m_t-m_s$ and $V_{\hbox{\scriptsize{dip}}}$ as shown in the right side of Fig.~\ref{fig4}. However, by extrapolating to $V_{\hbox{\scriptsize{dip}}}=0$~mV$_{\hbox{\scriptsize{pp}}}$, the resulting $(\xi _{\hbox{\scriptsize{exp}}}-\xi _{\hbox{\scriptsize{bib}}})/\xi _{\hbox{\scriptsize{bib}}}$ for $^{44}$Ca$^+$-$^{40}$Ca$^+$ and $^{48}$Ca$^+$-$^{40}$Ca$^+$, are of the same order as their uncertainties. It is important to note that if $V_{\hbox{\scriptsize{dip}}}\geq 4$~mV$_{\hbox{\scriptsize{pp}}}$, the ordered structure only remains for the \mbox{$^{42}$Ca$^+$-$^{40}$Ca$^+$} crystal (see Suppl. material). $z_s^{\hbox{\scriptsize{max}}}$, defined as the oscillation amplitude for $z_s^{\hbox{\scriptsize{exc}}}$ shown, e.g. in Fig.~\ref{fig3} for a \mbox{$^{48}$Ca$^+$ - $^{40}$Ca$^+$} crystal driven with $V_{\hbox{\scriptsize{dip}}}=2$~mV$_{\hbox{\scriptsize{pp}}}$, increases as a function of $V_{\hbox{\scriptsize{dip}}}$. For the \mbox{$^{42}$Ca$^+$ - $^{40}$Ca$^+$} crystal, the largest $z_s^{\hbox{\scriptsize{max}}}$ ranged from 5.70(3)~$\mu$m to 17.4(1)~$\mu$m, for the $V_{\hbox{\scriptsize{dip}}}$ values of Fig.~\ref{fig4}. These amplitudes are in most cases larger than the distance between the two ions in the unperturbed crystal, of about 8~$\mu$m.

The effect of the Coulomb interaction on increasing the oscillation amplitude of the ions in the crystal has been considered as a source for this systematic shift. Numerical calculations have been carried out considering only the axial modes and no influence of the radiofrequency trapping field. The systematic frequency shifts observed due to the excitation of the crystals are proportional to $z_s^2$, with a proportionality constant depending on the mass difference $C(m_t-m_s)$, assuming $q_s=q_t=1$. However, they are below those observed in the experiments and with opposite sign, which suggests a different reason than the Coulomb interaction or anharmonicities in symmetric potentials \cite{home2011}. The effect of the constant radiation-pressure force induced by the red-detuned laser has been also considered. This would lead to a displacement of the ions from their expected equilibrium positions by an amount $\delta z _{\hbox{\scriptsize{laser}}}$ that can shift the eigenfrequencies of the two-ion system proportionally in first order to $\delta z _{\hbox{\scriptsize{laser}}}/D$ \cite{staanum2004}. Frequency shifts in the order of $1 \times 10^{-5}$ or smaller have been obtained for the range of masses studied in this paper, negligible compared to the statistical uncertainty of the measurements.


In this paper we have demonstrated the non-destructive identification of sympathetically-cooled ions in a linear trap using a radiofrequency comb-like structure that excites the axial motion of the two-ion crystal. The method is fast and sufficient to discriminate between atomic species of the periodic table with a mass resolving power $m/\Delta m_\mathrm{FWHM}\approx 310$, providing at the same time a platform for single-ion laser spectroscopy. Single measurements within times below 1~s, can provide the mass identification in a regime with minimum disturbances due to the Coulomb interaction. 

We have implemented a sequence to measure the axial frequency of an individual laser-cooled ion (sensor ion), before and after the axial-common mode frequency of the unbalanced crystal, formed by the sensor and a `dark" target ion, to evaluate the mass identification method. When $m_t-m_s < 4$, the motional frequency ratios follow Eq.~(\ref{eq:mass_ratio_crystal}) with a relative uncertainty of $10^{-4}$. When $m_t-m_s \geq 4$ it deviates as a function of $m_t-m_s$ and of the oscillation amplitude of the ions when probing the crystal. The outcomes from simulations show that the effect of the Coulomb interaction on the axial modes is not visible in the measurements presented in this manuscript and the deviations observed when increasing the oscillation amplitudes of the ion in the crystal have to be attributed to other factor, still mass dependent, which introduces deviations from the values predicted by Eq.~(\ref{eq:mass_ratio_crystal}). This is important when using this method for larger differences between the mass-to-charge ratio of the target and sensor ions. Heavier ions will need other ion species which have similar Doppler cooling schemes than Ca$^+$, like $^{86,87,88}$Sr$^+$ or $^{134-138}$Ba$^+$ to measure SHEs (with $A\approx 258$) with charge states of 3$^+$ and 2$^+$, respectively, in order to keep $m_t/q_t-m_s/q_s$ smaller and thus, prevent large systematic shifts. 

If the conditions in the trap are highly stable, one single-ion identification can be performed prior to a laser-spectroscopy experiment or after e.g., resonance ionization spectroscopy.  Although our method is conceptually feasible to be applied to SHEs, the linear Paul trap should be placed in a beam line where the target ion can be introduced with high efficiency. This requires a cryogenic buffer-gas stopping cell \cite{kaleja2020}, to thermalize the ions and extract and guide them at very low energies, into a preparation trap such as an RFQ buncher to produce pulses for efficient injection in the linear Paul trap. The Penning trap facility SHIPTRAP has these elements \cite{block2005} and the coupling of such a system can be done after the RFQ buncher by means of electrostatic optics. It is worth mentioning that the results from the axial motion can be projected to a Penning trap \cite{gutierrez2019}, preventing possible perturbative effects due to the radiofrequency field. \\

\section*{Acknowledgement}

We acknowledge support from Grant No. PID2022-141496NB-I00 funded by MCIU/AEI /10.13039/501100011033 and by ERDF, EU, Grant No. PID2019-104093GB-I00 funded by MCIU/AEI /10.13039/501100011033, from FEDER/Junta de Andalucía - Consejería de Universidad, Investigación e Innovación through Project No. P18-FR-3432, from Programa ``Yo Investigo" Junta de Andaluc\'ia-Next Generation EU, and from the University of Granada ``Laboratorios Singulares 2020". The construction of the facility was supported by the European Research Council (Contract No. 278648-TRAPSENSOR), Projects No. FPA2015-67694-P (funded by MCIU/AEI/10.13039/501100011033 and by ERDF A way of making Europe) and No. FPA2012-32076 (MCIU/FEDER), infrastructure Projects No. UNGR10-1E-501, and No. UNGR13-1E-1830 (MCIU/FEDER/UGR), and No. EQC2018-005130-P (funded by MCIU/AEI/10.13039/501100011033 and by ERDF A way of making Europe), and infrastructure Projects No. INF-2011-57131 and No. IE2017-5513 (funded by Junta de Andalucía/FEDER).


%

\end{document}